\documentclass[ nobibnotes, nofootinbib,
preprint,
showpacs,
preprintnumbers,
amsmath,
amssymb]{revtex4-1}

\usepackage{graphicx}
\usepackage{dcolumn}
\usepackage{bm}
\usepackage{amssymb,wasysym}
\usepackage{hyperref}
\usepackage{mathrsfs}
\usepackage{color}

\def\b{{\textsc{b}}}
\def\a{{\textsc{a}}}
\def\d{{\textsc{d}}}

\begin{document}


\title{Meissner effect and a stringlike interaction}
\author{Chandrasekhar Chatterjee}
\email{chandra@phys-h.keio.ac.jp, chandra.chttrj@gmail.com}
\affiliation{$^1$ Department of Physics at Hiyoshi, and Research and Education Center for Natural Sciences,\\ Keio University,Hiyoshi 4-1-1, Yokohama, Kanagawa 223-8521, Japan}
\author{Ishita Dutta Choudhury}
\email{ishitadutta.choudhury@bose.res.in}
\affiliation{S N Bose National Centre for Basic Sciences, Block JD, Sector III, Salt Lake, Kolkata 700106, India}
\author{Amitabha Lahiri}
\email{amitabha@bose.res.in}
\affiliation{S N Bose National Centre for Basic Sciences, Block JD, Sector III, Salt Lake, Kolkata 700106, India}

%
%


\date{\today}

\begin{abstract}
We find that a recently proposed interaction involving the vorticity current of electrons, 
which radiatively induces a photon mass in 3+1 dimensions in the low-energy effective 
theory, corresponds to confining strings (linear potential) between electrons. 
\end{abstract}
\maketitle

The formal theory of superconductivity was developed on the principle of 
Cooper pair formation~\cite{Cooper:1956zz,Bardeen:1957mv,Bogolyubov:1958se}. 
According to this principle, the interaction between electrons and phonons generate 
an effective electron-electron interaction limited to a shell in momentum space around the Fermi surface. 
When this mutually attractive interaction overcomes the Coulomb repulsion, 
bound pairs may form. In conventional superconductivity, the bound pair of 
electrons which possess mutually opposite momenta are in an s-wave state 
(spin singlet)~\cite{Cooper:1956zz,Bardeen:1957mv,Bogolyubov:1958se}. 
In this case, the theory has a local order parameter of spin zero and 
the system undergoes spontaneous symmetry breaking. However, there are
superconductors which do not seem to exactly follow the above description.
For instance, in the case of  topological superconductor, the system 
may not have any local order parameter  like in the 
Ginzburg-Landau description~\cite{Hansson:2004wca,Diamantini:2014iqa}.
Furthermore, in an unconventional superconductor the pair states 
can have non-zero spin. Although a spinorial order parameter was first 
discovered  in superfluid He-3~\cite{Osheroff:1972}, unconventional 
superconductivity can be realized in many heavy fermionic compounds. 
It is well known today that spin interactions, and particularly spin-orbit coupling, play
an important role in the physics of topological matter~\cite{Galitski:2013iia}. 
It has also been shown that long range spin-spin interactions can be induced  by
collective excitations, like phonons~\cite{Bennet:2013}.

As is well known, the expulsion of magnetic field from superconductors, or Meissner effect, is described by 
the London equations, 
\begin{equation}
\partial_\mu J_\nu - \partial_\nu J_\mu = \lambda\, F_{\mu\nu}\,.
\label{London}
\end{equation}
In conventional superconductors, this is implemented by an underlying model of 
symmetry breaking, in which a complex Higgs field $H$ of charge $2e$ acquires 
a vacuum expectation value $v$\,. Then the azimuthal part of $H$, which may 
be written as $e^{i\phi}$ and is the Nambu-Goldstone mode of the 
symmetry breaking, is responsible for a current
\begin{equation}
J_\mu = 4i\, e v^2 \, e^{-i\phi} D_\mu e^{i\phi}\,, \qquad  D_\mu = \partial_\mu - 2ie\,A_\mu\,.
\end{equation}
This current is gauge invariant, and reduces the London equation to an identity, with 
$\lambda = 8e^2 v^2\,.$ 

It is by now well known that there is a `dual' ansatz for the current which does not require 
symmetry breaking. Instead of a complex scalar field with a vacuum expectation value, an 
antisymmetric tensor field $B_{\mu\nu}$ is introduced, with field strength $H_{\mu\nu\lambda} 
= \partial_{\mu}B_{\nu\lambda} + \partial_\nu B_{\lambda\mu} + \partial_\lambda B_{\mu\nu}\,,$ \cite{Kalb:1974yc}
and the current is defined to be $J^\mu = \epsilon^{\mu\nu\lambda\rho} H_{\nu\lambda\rho}\,.$
Then the continuity equation $\partial_\mu J^\mu = 0$ becomes an identity, and Eq.~(\ref{London})
can be derived as field equations from the Lagrangian~\cite{Cremmer:1973mg,Allen:1990gb}
\begin{equation}
	S =  \int d^{4}x\,\, \left(- \frac{1}{4}\, F_{\mu \nu}F^{\mu \nu} 
	+ \frac{1}{12}\, H_{\mu \nu \alpha} H^{\mu \nu \alpha} 
	+ \frac{m}{4}\,\epsilon^{\mu\nu\lambda\rho} B_{\mu\nu} F_{\lambda\rho} 
	\right)\,,
	\label{bf.abl}
\end{equation}
with $\lambda = 2m^2\,.$ This is usually referred to as a topological mass generation
mechanism for the photon because of the $B\wedge F$ interaction term~\cite{Bergeron:1994ym}, 
which by itself 
is an Abelian version of a similar term in many four dimensional topological quantum 
field theories. This interaction is a four dimensional generalization of the Chern-Simons 
term in three dimensions~\cite{Jackiw:1980kv,Deser:1982vy,Deser:1981wh}, but unlike the 
latter it does not break $P$ and $T$.

Chern-Simons theory finds application in three dimensional condensed matter physics, 
in particular in describing fractional quantum Hall effect~\cite{Wen:2012hm}. The action
corresponding to Eq.~(\ref{bf.abl}) is a mixed Chern-Simons theory in three dimensions, 
and has been used in quantum hall systems and in superconductors with a gap
in the single particle spectrum~\cite{Hansson:2004wca}. This description relies on the
duality between the $BF$ action of Eq.~(\ref{bf.abl}) and the Abelian Higgs model, albeit
with a frozen Higgs degree of freedom. In four dimensions however, there is a subtle 
distinction between the two dual theories, because $B_{\mu\nu}$ can in principle have 
couplings which have no analogue in the Higgs picture. It has been argued 
in~\cite{Diamantini:2014iqa} that this implies that superconductors described by
$BF$ theory are different from the usual kind. Indeed, the four dimensional theory 
of Eq.~(\ref{bf.abl}) has proven very difficult to implement in realistic systems. 
This difficulty is based on the lack of a sensible interaction 
between fermions and the $B_{\mu\nu}$ field, namely, one that is invariant under the vector gauge 
transformation $B_{\mu\nu} \to B_{\mu\nu} + \partial_\mu \chi_\nu - \partial_\nu \chi_\mu\,.$
It was shown recently that such an interaction could be constructed, provided $B_{\mu\nu}$
was coupled to a non-local fermion current. Such a coupling induces the $B\wedge F$ term
at one loop level even if it is not present in the original Lagrangian~\cite{Choudhury:2015rua}, 
and the pseudovector charge density of the non-local current can be interpreted as the vorticity 
field of the fermion. In other words, a mass for the photon is induced by the non-local interactions,
producing an alternative low-energy effective theory of superconductivity, valid for 
energy scales well below an ultraviolet cut-off $\Lambda$. The induced photon mass 
is cut-off dependent, being proportional to $m\log \frac{\Lambda^2}{m^2}$, with 
$m$ the fermion mass. 

The form of the coupling between fermions and the $B_{\mu\nu}$ field raises 
an interesting question.
Just as the ordinary gauge potential couples to worldlines of charged particles, the 
antisymmetric tensor gauge field $B_{\mu\nu}$ couples naturally to worldsheets swept out by strings.
Interactions between strings can be thought of as being mediated by the $B$ field, just as 
interactions between charged particles can be thought of as being mediated by vector gauge 
fields, as has been known from the early days of string theory. Now that we have a coupling 
between the $B$ field and fermions, a coupling which is not localized at a point, the question 
naturally arises as to whether the system contains stringlike objects, and what 
such strings would  be made of. 

In this letter we derive the remarkable result that the static potential between a
pair of fermions in this theory has a component that is linear and attractive, independent 
of the charge of  the fermions. For a pair of electrons, this is like a Cooper pair connected by 
a confining string.

The result is fairly easy to obtain. We start with the partition function
\begin{equation}
	Z= \int {\mathscr D} B \,{\mathscr D}A\, {\mathscr D} \bar{\psi}\, {\mathscr D} \psi\, e^{iS}
\end{equation}
for the action~\cite{Choudhury:2015rua}
%
%
%
\begin{eqnarray}
	S = \int d^{4}x &\left[ - \frac{1}{4} F_{\mu \nu}F^{\mu \nu} 
	+ \frac{1}{12} H_{\mu \nu \lambda} H^{\mu \nu \lambda}  
	+\bar{\psi}(i \gamma_{\mu}\partial^{\mu}-m)\psi 
	+ g\,m\, \epsilon^{\mu \nu \lambda\sigma}\,B_{\mu \nu} 
	\frac{\partial_{\lambda}}{\square}\, 
	\bar{\psi} \gamma_{\sigma} \psi
	 + eA_{\mu}\bar{\psi} \gamma^{\mu} \psi
	\right]\, .
	\label{action}
\end{eqnarray}
Here $\psi(x)$ is a charged fermion of mass $m$ interacting with electromagnetic gauge field $A_\mu$ with coupling $e\,,$ and also with  an anti-symmetric tensor field $B_{\mu\nu}$\, with coupling $g$.
This action is invariant under the aforesaid vector gauge symmetry in addition to the usual 
$U(1)$ gauge symmetry. Therefore, in order to perform the integrals over $B_{\mu \nu}$ and 
$A_{\mu}$, we add the gauge fixing terms $-\frac{1}{2\zeta}(\partial_{\mu} B^{\mu \nu})^2 
 -\frac{1}{2\eta}(\partial_{\mu} A^{\mu})^2 $ to the Lagrangian.

Integrating over the $A$ and $B$ fields, we get 
\begin{equation}
	Z = N \int {\cal D}\bar{\psi}\, {\cal D}\psi\,\, e^{iS[\psi\,, \bar{\psi}]}\, ,
\end{equation}
where
%
\begin{eqnarray}
	S[\psi,\bar{\psi} ] = & \,\,S_\d[\psi, m] 
	+ \frac{1}{2} \int  \frac{d^4k}{(2\pi)^4} \,\, 
	\left[J^{\sigma}(-k)\, \frac{e^2}{k^2}\,J_{\sigma}(k)
	- {g^2} J_{\mu \nu}(-k)\, \frac{g^{\mu[\rho}\,g^{\lambda ]\nu}}{k^2}\, J_{\rho \lambda}(k)\right]\, \nonumber \\
	= &  S_\d [\psi, m] \,+\,  S_\a[\psi,\bar{\psi} ] \,+\,  S_\b[\psi,\bar{\psi} ] \,. 
	\label{eff.act}
\end{eqnarray}
$S_\d[\psi, m]\,$ consists of the kinetic and mass terms of the fermion, and  in writing 
the other two terms we have used the fact that both $J_\mu$ and $J_{\mu\nu}$ are conserved currents.

The antisymmetric tensor current $J_{\mu\nu}$ is defined to be what couples to $B_{\mu\nu}$ 
in Eq.~(\ref{action}), and it is easy to see that 
it is related to the fermion current $J_\mu$ by
\begin{eqnarray}
	J_{\mu \nu}(k) 
	=  -\,m\,\epsilon_{\alpha \mu \nu \sigma} \frac{ik^\alpha}{k^2} J^{\sigma}(k)\,,
	\label{current}
\end{eqnarray}
which when inserted into the last term of Eq.~(\ref{eff.act}) gives 
\begin{eqnarray}
	S_\b[\psi,\bar{\psi} ] = 2 \int \frac{d^4k}{(2\pi)^4} \,J^{\sigma}(-k)\,\frac{g^2 m^2}{k^4}\,J_{\sigma}(k)\,.
\end{eqnarray}

To obtain the form of the potential in non-relativistic limit, we expand the fermion 
fields in terms of annihilation and creation operators $a_r(\bf{p})$ and
$a^\dagger_r(\bf{p})$ etc. We note that in the non-relativistic limit, only 
the $J^0$ component will contribute to the effective action of Eq.~(\ref{eff.act}) because 
$J^i$ is made of the lower components of Dirac spinors and thus can be neglected for 
energies much lower than their mass. Therefore, the leading contribution from 
this term is
\begin{equation}
	\int \frac{d^3p}{(2\pi)^3} \frac{d^3p^{\prime}}{(2\pi)^3} \frac{d^3q}{(2\pi)^3} \frac{d^3q^\prime}{(2\pi)^3} 
	\sum_{r, s} a^{\dagger}_{r}
	({\bf q^{\prime}})a_{r}({\bf q}) \left[\frac{e^2}{(p-p^{\prime})^2} + 2 \frac{g^2 m^2}{(p-p^{\prime})^4}\right]
	a^{\dagger}_{s}({\bf p^{\prime}})a_{s}({\bf p})\,.
	\label{scat.amp}
\end{equation}

In the non-relativistic limit, we keep terms only to the lowest order in the 3-momenta  so 
that we can write $	p= ( m, {\bf p})\,,\,
p^{\prime} = ( m, {\bf p^{\prime}})\,,$
and
\begin{equation}
	(p-p^{\prime})^2 \approx - \vert {\bf p}-{\bf p^{\prime}} \vert ^2 \,.
\end{equation}
What we have calculated in Eq.~(\ref{scat.amp}) is an effective scattering amplitude for electron-electron 
interaction. The static potential has an extra factor of $(-1)$ compared to this~\cite{Peskin:1995ev},  which means that in three 
dimensional momentum space the static potential has the form
\begin{equation}
	V({\bf k}) = \frac{e^2}{|{\bf k}|^2} - 2 \frac{g^2 m^2}{|{\bf k}|^4}\,,
\end{equation}  
where ${\bf k} = {\bf p} - {\bf p'}$\,.

The Fourier transform of $V(\textbf{k})$ gives the expression for the static potential in 
three dimensional coordinate space,
\begin{eqnarray}
	V(r) = \frac{e^2}{4\pi r} + \frac{g^2 m^2 r }{2 \pi}\,.
	\label{stat.pot}
\end{eqnarray}
The force resulting from the static potential given in Eq.~(\ref{stat.pot}) has the form
\begin{equation}
	- {\bf \nabla} V(r) = \left(\frac{e^2}{4\pi r^2} - \frac{g^2 m^2 }{2 \pi}\right) \hat{\bf r}\,.
	\label{force}
\end{equation}

The first term in Eq.~(\ref{force}) is the Coulomb force which is repulsive for the electron-electron interaction. 
But the second term, which comes from a confining linear potential and is independent 
of the electric charges of the fermions, is attractive for the electron-electron interaction. 
Following the idea of ``Cooper instability''~\cite{Cooper:1956zz,Bardeen:1957mv}, 
we can say that the attractive interaction generates a kind of pairing between electrons.
For chargeless fermions, this pairing 
would be responsible for  gauge-invariant mass generation of $B_{\mu\nu}$ similar to Schwinger 
mechanism~\cite{Schwinger:1962tn,Schwinger:1962tp}\,, by shifting the pole of the $B_{\mu\nu}$ 
field. When the fermions couple also to $A_\mu$, it  generates an effective $B\wedge F$ 
interaction. This shows that although the fermions are confined by linear confining 
potential, the interacting field  $B_{\mu\nu}$ is not confined, rather it would behave 
as a short-ranged force field.

    The action in eq.~(\ref{action}) we started with  describes low-energy effective interactions of the system. So the exact 
description of the interactions at very short distance scale may not be possible. 
However, we may make an estimation of the scale below which photons become short ranged. 
This can be thought of as the separation of the pair, or a `correlation length' $r_{\rm cor}$, which can 
be estimated by setting the net force to zero,
\begin{equation}
	\frac{1}{{r^2}_{\rm cor}} = \frac{{2}g^2 m^2}{e^2}\,{.}
	\label{corr.length}
\end{equation}
At ${r}_{\rm cor}$, the potential has the value $\frac{egm}{2{\sqrt{2}}\pi}$\,.

We may think of the potential of Eq.~(\ref{stat.pot}) as describing a system of two spins (electrons) 
connected by a string of fixed length. If the length of the string exceeds the value given in Eq.~(\ref{corr.length}), 
the attractive linear potential dominates, and the spins are attracted to each other. If they come closer than 
${r}_{\rm cor}$, the Coulomb repulsion dominates, leading to the spins moving away from each other. 
We can also expect that the effect of interactions with other nearby spins could modify the structure of the 
string.  As it is clear from Eq~(\ref{scat.amp}) the attractive interaction is independent of spin, so the {Cooper 
pair wave function}  can have degenerate spin combinations 
\begin{eqnarray}
C_{1}\,|\uparrow, \,\uparrow\rangle + C_{01}\, |\uparrow, \,\downarrow\rangle + C_{02}\, |\downarrow,\, \uparrow\rangle + C_{-1} \, |\downarrow, \,\downarrow\rangle.
\label{pairs}
\end{eqnarray}
{Here the coefficients in general are  functions of space time coordinates and orbital angular momentum.}
This indicates the system can be in a spin one or spin zero superconducting phase.

How does our work relate to, or differ from, earlier work? It is obvious that the confining
potential of Eq.~(\ref{stat.pot}) must appear when the interacting charged particles 
are connected by a flux tube or string. The Kalb-Ramond field $B_{\mu\nu}$ also arises 
naturally in such situations. For example, in the context of confining strings in gauge
theories~\cite{Polyakov:1996nc,Diamantini:1996vf,Quevedo:1996uu}, or in a field
theoretic description of topological matter~\cite{Diamantini:2005dj,Diamantini:2011ck}, the 
antisymmetric tensor field $B_{\mu\nu}$ originates from the condensation 
of topological charge in compact U(1) gauge theory. In these cases one finds flux tubes
carrying electric (or magnetic, or both electric and magnetic) flux. The model we have considered
differs from these in both respects. The $B$ field, and its non-local coupling with fermions
which was our starting point in this investigation, must arise from some underlying theory 
in terms of an effective action, but the details of that process is not clear. More importantly,
it is not clear what corresponds to flux tubes in this model, since electric charge is unconfined,
and what couples to the $B$ field does not appear to be magnetic charge. Further, the
$B$ interaction does not appear to give rise semi-classically to topological defects. 
While the current which couples to $B_{\mu\nu}$ corresponds to the vorticity of the
fermion field in a semi-classical picture, this does not lead to a winding number for a given pair 
of fermions. The effective potential of Eq.~(\ref{stat.pot}) comes from a purely quantum calculation, 
and in that regard our results are perhaps closest in spirit to the ideas of~\cite{'tHooft:2002wz}.
 
 To summarize, in this letter we have described a model which exhibits an unconventional 
pairing of fermions, and at the same time 
produces Meissner effect at low energy by a radiatively induced effective 
$ B\wedge F$ interaction. This process does not involve any spontaneous 
symmetry breaking  by any local order parameter~\cite{Hansson:2004wca,Norman:2013}. It could be useful as a description of unconventional 
superconductors which expels magnetic field from the bulk but pairs fermions
by flux tubes or something analogous. 
The main result of our paper is that the effective 
potential between two pairing electrons has a part which is linear and attractive, 
which means that the theory is in a confining phase. 
 According to the 
't\,Hooft-Mandelstam~\cite{Mandelstam:1974pi,'tHooft:1975pu} 
description of confinement, the system must be in 
a disordered state~\cite{'tHooft:1977hy}. So the non-existence of local 
order parameter is natural here, although following order-disorder 
duality, it may be possible to describe the confining system by a dual 
order parameter. In the present case we have not found such 
a dual order parameter yet, but Eq.~(\ref{pairs}) shows that pair
formation due to the confinement mechanism would create a kind 
of ordering.




\section*{Acknowledgements}
This work is partially supported by the Ministry of Education, Culture, Sports, Science (MEXT)-Supported Program for the Strategic Research Foundation at Private Universities ``Topological Science'' (Grant No. S1511006).
 C.~C. acknowledges support as an overseas researcher under Postdoctoral Fellowship of Japan Society for the Promotion of Science(JSPS).

\end{document}